\documentclass{aa}
\usepackage{amsmath,graphicx,amssymb}
\usepackage[varg]{txfonts}
\usepackage{natbib}
\usepackage{pst-eps}

\newcommand{\zav}[1]{\left(#1\right)}

\newlength\staretab
\newcommand{\Teff}{\ensuremath{T_\mathrm{eff}}}
\newcommand{\logg}{\ensuremath{\log (g/1\,\text{cm}\,\text{s}^{-2})}}

\newcommand\de{\text{d}}
\newcommand\x[1]{\ensuremath{#1_\text{X}}}
\newcommand\lx{\ensuremath{\x L}}

\newcommand\msr{\ensuremath{M_\odot\,\text{yr}^{-1}}}
\newcommand\kms{\ensuremath{\text{km}\,\text{s}^{-1}}}

\newcommand\diag{$T_\text{eff}$ versus $\log g$}
\newcommand\ergs{\ensuremath{\text{erg}\,\text{s}^{-1}}}

\newcommand\radec[6]{$\alpha={#1\,\text{h}\,#2\;\text{m}\;#3\,\text{s},}\;\delta={#4^\circ\,#5'\,#6''}$, J2000}

\begin{document}

\title{Hot subdwarf wind models with accurate abundances}
\subtitle{I. Hydrogen dominated stars HD 49798 and
BD+18$^\circ$\,2647\thanks{Based on observations collected at the European
Southern Observatory, Paranal, Chile (ESO programme 097.D-0540(A)).}}

\author{J.~Krti\v{c}ka\inst{1} \and J.~Jan\'\i k\inst{1} \and
I.~Krti\v{c}kov\'a\inst{1} \and S.~Mereghetti\inst{2} \and F.~Pintore\inst{2}
\and P.~N\'emeth\inst{3} \and J.~Kub\'at\inst{3} \and M.~Vu\v ckovi\'c\inst{4}}

\institute{\'Ustav teoretick\'e fyziky a astrofyziky, Masarykova univerzita,
           Kotl\'a\v rsk\' a 2, CZ-611\,37 Brno, Czech
           Republic
           \and
           Istituto di Astrofisica Spaziale e Fisica Cosmica Milano,
           via Alfonso Corti 12, 20133, Milano, Italy 
           \and
           Astronomick\'y \'ustav, Akademie v\v{e}d \v{C}esk\'e
           republiky, Fri\v{c}ova 298, CZ-251 65 Ond\v{r}ejov, Czech Republic
           \and
           Instituto de F\'\i sica y Astronom\'\i a, Facultad de Ciencias,
           Universidad de Valpara\'\i so, Gran Breta\~na 1111, Playa Ancha,
           2360102, Valpara\'\i so, Chile 
}

\date{Received}

\abstract{Hot subdwarfs are helium burning objects in late stages of their
evolution. These subluminous stars can develop winds driven by light absorption
in the lines of heavier elements. The wind strength depends on chemical
composition which can significantly vary from star to star.} {We aim to
understand the influence of metallicity on the strength of the winds of the hot
hydrogen-rich subdwarfs HD 49798 and BD+18$^\circ$\,2647.} {We used
high-resolution UV and optical spectra to derive stellar parameters and
abundances using the TLUSTY and SYNSPEC codes. For derived stellar parameters,
we predicted wind structure (including mass-loss rates and terminal velocities)
with our METUJE code.} {We derived effective temperature
$T_\text{eff}=45\,900\,$K and mass $M=1.46\,M_\odot$ for HD 49798 and
$T_\text{eff}=73\,000\,$K and $M=0.38\,M_\odot$ for BD+18$^\circ$\,2647. The
derived surface abundances can be interpreted as a result of interplay between
stellar evolution and diffusion. The subdwarf HD 49798 has a strong wind that does not allow
for chemical separation and consequently the star shows solar chemical
composition modified by hydrogen burning. On the other hand, we did not find
any wind in BD+18$^\circ$\,2647 and its abundances are therefore most likely
affected by radiative diffusion. Accurate abundances do not lead to a
significant modification of wind mass-loss rate for HD 49798, because the
increase of the contribution of iron and nickel to the radiative force is
compensated by the decrease of the radiative force due to other elements. The
resulting wind mass-loss rate $\dot M=2.1\times10^{-9}\,\msr$ predicts an X-ray
light curve during the eclipse which closely agrees with observations. On the
other hand, the absence of the wind in BD+18$^\circ$\,2647 for accurate
abundances is a result of its peculiar chemical composition.} {Wind models with
accurate abundances provide more reliable wind parameters, but the influence of
abundances on the wind parameters is limited in many cases.}


\titlerunning{Hot subdwarf wind models with accurate abundances: I.~HD 49798 and
BD+18$^\circ$\,2647}

\maketitle

\section{Introduction}

The stellar winds of hot stars are driven by the radiative force due to light
absorption in the lines of heavy elements \citep{cak,pulvina}. Consequently,
metallicity, in addition to stellar luminosity, is one of the key parameters that
determine the properties of hot star winds. Thanks to a relatively low
metallicity gradient in our Galaxy \citep[e.g.][]{martinzgrad}, wind studies of
Galactic main sequence and supergiant massive stars may safely assume solar
chemical composition in most cases. Moreover, although the mixing induced by
stellar rotation may alter the surface chemical composition during stellar
evolution \citep{memarot}, corresponding variations of wind parameters are
typically negligible for solar metallicity stars whose surfaces are enriched by
hydrogen burning products \citep{dusik}. This further justifies the assumption
of solar metallicity for studies of main sequence and supergiant hot stars.

With the advent of 8m class telescopes and the Hubble Space Telescope (HST) it
became possible to spectroscopically study winds from individual hot stars
residing in the Local Group of galaxies \citep[e.g.][]{mamrac2,nemracna,sabina}.
Although these stars have non-solar abundances, a simple assumption of scaled
solar chemical composition seems to be sufficient for the study of their winds
\citep{vikolamet,mcmfkont}. 

However, the assumption of scaled solar chemical composition drops at late
evolutionary phases of massive stars during the Wolf-Rayet phase when the
envelope is stripped in the course of single or binary star evolution
\citep{vanvanwr,sylsit}. In these late evolutionary phases, products of hydrogen
and helium burning appear on the stellar surface. This results in strong
deviations from solar chemical composition, which has severe consequences for
mass-loss \citep{grahamz}.

\begin{table*}[t]
\caption{Spectra used for the analysis.}
\centering
\label{pozor}
\begin{tabular}{lcccc}
\hline
Star  & Instrument & Spectrum & Domain [\AA] & JD$-2\,400\,000$ \\
\hline
HD 49798
& UVES@UT2 & UVES.2016-10-02T09:03:11.546 & 3732--5000 & 57663.87826\\
& HST/GHRS & z0x60606t & 1248--1270 & 48733.94266\\
&& z0x60607t & 1300--1335 & 48733.94516\\
&& z0x60608t & 1602--1637 & 48733.94818\\
&& z0x60609t & 1650--1684 & 48733.95195\\
&& z0x6060at & 1800--1834 & 48733.95564\\
&& z0x6060ct & 1840--1874 & 48733.97662\\
\hline
BD+18$^\circ$\,2647 
& UVES@UT2 & UVES.2016-07-29T23:44:26.000 & 3732--5000 & 57599.49729\\
& IUE/SWP & 20488 & 1150--1975 & 45536.48785\\
& FUSE & m1080701000 & 900--1190 & 51664.15527 \\
\hline
\end{tabular}
\end{table*}

The effects of deviations from solar chemical composition on stellar winds can also
be expected in hot subdwarfs. These stars, stripped of their envelope, are
low-mass counterparts to Wolf-Rayet stars \citep{bozihrad} and therefore also
show non-solar chemical composition. Moreover, the processes of radiative
diffusion and gravitational settling may affect the surface chemical composition
of subdwarfs \citep{unbu,vanrimi,miriri,hu}.

Subdwarfs share with Wolf-Rayet stars not only non-solar chemical composition
but probably also similar origin. There are several possible evolutionary
channels that can lead to the appearance of subdwarfs. Helium low-luminosity
subdwarfs may originate as a result of merging of two white dwarfs
\citep{iba,saje,zhaff} or due to helium core flashes while descending on the white
dwarf cooling track directly after the departure from the red giant branch
\citep[`hot flasher' scenario, e.g.][]{brflasher,batika}. Subluminous stars may also
be products of red giants stripped of their envelopes most
likely during their binary evolution \citep[e.g.][]{jinanvoxfordu}.

Despite their likely non-solar chemical composition, stellar winds of hot
subdwarfs were studied assuming either solar or scaled solar chemical
composition \citep{vinca,un,snehurka}. This is likely not the most suitable
approach to late evolutionary stages, when the assumption of scaled solar
composition does not provide a precise estimate of wind structure. Moreover, some
hot subdwarfs emit X-rays, which are supposed to originate in their winds
\citep{bufacek}. The non-solar chemical composition could explain why some hot
subdwarfs are located far away from the canonical relationship between X-ray
luminosity and bolometric luminosity \citep{snehurka}.

To understand the role of non-solar chemical composition, we initiated an
observing campaign, during which we plan to derive detailed photospheric
properties of selected hot subdwarfs and to simulate their winds with accurate
chemical composition. We selected subdwarfs that emit lower amounts of X-rays
than expected from the mean relationship between X-ray luminosity and bolometric
luminosity \citep[Fig.~5]{snehurka}. Here we present the results of such a study
for two hydrogen-dominated subdwarfs.

\section{Spectroscopy} 

The spectral analysis presented here of hydrogen-dominated subdwarfs is based on our own
optical spectroscopy and on archival ultraviolet (UV) data. We obtained
observational time with the high-resolution spectrograph UVES $(R=80\,000)$
located at the Nasmyth B focus of VLT-UT2 (Kueyen) via ESO proposal
097.D-0540(A). The spectrum of the star BD+18$^\circ$\,2647 was taken on
29\,July,\,2016, with a total integration time of $1400$s, and of the star HD~49798
on 2\,October,\,2016, with an exposure time of 180s. For reduction, we used standard
IRAF\footnote{IRAF is distributed by NOAO, which is operated by AURA, Inc.,
under cooperative agreement with the National Science Foundation.} routines
(bias, flat, and wavelength calibration). Both spectra cover the spectral region
of $3732 - 5000~\AA$. 

In the UV domain we used spectra obtained by IUE (SWP camera, high-dispersion data),
FUSE, and HST (GHRS spectrograph) satellites. The processed spectra were
downloaded from the MAST archive\footnote{Mikulski Archive for Space Telescopes,
http://archive.stsci.edu}. The list of all used observations is given in
Table~\ref{pozor}.

\section{Analysis of spectra}

The spectroscopic analysis of subdwarf spectra was based on the hydrostatic NLTE
model atmosphere code TLUSTY \citep{tlusty0} version 200. The hydrostatic
model atmospheres give reliable stellar parameters even for stars with winds
\citep{bourak,kupa}, provided the mass-loss rates are low, as in the case of
subdwarfs. The atomic data used for the atmosphere modelling are the same as in
\citet{ostar2003}. The data were mostly calculated within the Opacity and Iron
Projects \citep{topt,zel0}. Synthetic spectra were calculated from the model
atmospheres using the SYNSPEC code \citep{synspec} version 45. We also
measured the radial velocity from each UVES spectrum by means of a
cross-correlation function using the theoretical spectrum as a template
\citep{zvezimi}. 

The stellar parameters were determined using the $\chi^2$ minimisation of the
difference between observed and predicted spectra using the simplex
method \citep{kobr}. We derived stellar effective temperature $T_\text{eff}$,
surface gravity $\log g$, and abundances of individual elements
$\varepsilon_\text{el}$ for each star. The elemental abundances are given as
number density ratios relative to hydrogen, that is,
$\varepsilon_\text{el}={N_\text{el}/N_\text{H}}$.  The minimisation proceeded
in three steps:
\begin{enumerate}
\item We calculated a model atmosphere and a synthetic spectrum grid in
$T_\text{eff}$, $\log g$, and $\varepsilon_\text{He}$. At the beginning of
iterations we used a relatively broad range of stellar parameters, which was
subsequently made narrower as the parameters approached the final value. We
assumed a fixed number density ratio $N_\text{el}/\zav{N_\text{H}+N_\text{He}}$ of
heavier elements with respect to hydrogen and helium for elements whose
abundances were derived from spectra. For other elements we used solar
\citep{asp09} density ratio
$m_\text{el}N_\text{el}/\zav{m_\text{H}N_\text{H}+m_\text{He}N_\text{He}}$,
where $m_\text{el}$ is the atomic mass, assuming that the abundance of
these elements was not affected by nuclear reactions or by diffusion.
\item We minimised the $\chi^2$ differences between the observed spectrum and
the predicted spectrum interpolated from the grid to derive $T_\text{eff}$, $\log
g$, and $\varepsilon_\text{He}$.
\item We calculated a model atmosphere with $T_\text{eff}$, $\log
g$, and $\varepsilon_\text{He}$ derived in step 2 and based on this model
atmosphere we minimised the $\chi^2$ differences between the observed spectrum
and the predicted spectrum calculated for actual abundances of heavier elements.
\end{enumerate}
The steps 1 -- 3 were repeated until the changes of
parameters were lower than 1\%. The derived parameters are given in
Table~\ref{hvezpar}. The uncertainties on $T_\text{eff}$, $\log g$, and
$\varepsilon_\text{He}$ were estimated from fits of individual H and He lines.
The uncertainties of other elemental abundances were estimated from the
abundances derived from individual spectral regions.

We encountered some numerical difficulties during the computation of model
atmospheres. To resolve these difficulties, we followed general recommendations
\citep{tlusty3}; that is, we typically started with the LTE model and treated
lower levels assuming detailed radiative balance (using the ILVLIN parameter).
The model becomes more realistic with lower values of ILVLIN. It is usually
effective to start with high values of ILVLIN=100 for the models that have
difficulties in converging. After the successful calculation of the model with
a high value of ILVLIN, we used the result as an input for a following modelling and
progressively decreased the value of this parameter in subsequent steps. In some
cases, we successively decreased this parameter separately for individual
elements to obtain a pure NLTE model. We gradually included individual elements to
calculate more detailed models. Moreover, in some cases during the calculation
of an intermediate NLTE model we fixed the temperature in the outer layers to
achieve convergence. We relaxed this assumption during the calculation of
final NLTE models.

\section{Wind modelling}

We used the global wind code METUJE for the prediction of wind parameters
\citep{cmfkont}. METUJE provides global (unified) models of the stellar
photosphere and radiatively driven wind. The code solves the comoving frame
(CMF) radiative transfer equation, the kinetic (statistical) equilibrium
equations (often denoted as NLTE equations), and hydrodynamic equations from
an almost hydrostatic photosphere to a supersonically expanding wind. The
hydrodynamical equations contain the CMF radiative force calculated using NLTE
level populations. Therefore, the code predicts basic wind parameters including
the mass-loss rates $\dot M$ and the terminal velocities $v_\infty$ simply from
the stellar parameters. The code assumes a stationary (time-independent) and
spherically symmetric wind.

We calculated wind models for the stellar parameters ($\Teff$, $R$, and $M$)
given in Table~\ref{hvezpar} for both  subdwarfs studied here. To understand the role
of specific chemical composition in the driving of wind, we calculated two sets of
wind models. One set was calculated for derived stellar chemical
composition (yielding $\dot M$ and $v_\infty$) and the second set for solar
\citep{asp09} chemical composition (giving $\dot M^\odot$ and $v_\infty^\odot$).
These values are provided in Table~\ref{hvezpar}. We assumed solar chemical
composition for the elements whose abundances were not derived from spectra.

\section{HD 49798}
\newcommand\hvezda{\object{HD~49798}}

\begin{table}[t]
\caption{Derived parameters of studied stars.}
\label{hvezpar}
\begin{tabular}{lccc}
\hline
Parameter               & HD 49798         &  BD+18$^\circ$\,2647  & Sun \\
\hline
\Teff\ [K]              & $45\,900\pm800$  &  $73\,000\pm2000$\\
\logg                   & $4.56\pm0.08$    &  $5.95\pm0.03$ \\
$R$ [$R_\odot$]         & $1.05\pm0.06$    &  $0.107\pm0.011$\\
$M$ [$M_\odot$]         & $1.46\pm0.32$    &  $0.38\pm0.08$ \\
$\varepsilon_\text{He}$ & $0.74\pm0.07$    &  $0.029\pm0.010$ & $0.085$\\
$\log\varepsilon_\text{C}$  & $<-4.2$      &  $<-7.0$     & $ -3.57$ \\
$\log\varepsilon_\text{N}$  & $-3.1\pm0.2$ & $-4.5\pm0.5$ & $ -4.17$ \\
$\log\varepsilon_\text{O}$  & $-4.6\pm0.2$ & $-5.4\pm0.5$ & $ -3.31$ \\
$\log\varepsilon_\text{F}$  &              &  $<-8.0$     & $ -7.44$ \\
$\log\varepsilon_\text{Ne}$ &              &  $<-7.5$     & $ -4.07$ \\
$\log\varepsilon_\text{Mg}$ & $-4.2$                 &    & $ -4.40$ \\
$\log\varepsilon_\text{Al}$ & $-4.9\pm0.4$           &    & $ -5.55$ \\
$\log\varepsilon_\text{Si}$ & $-4.1\pm0.2$ & $-6.4\pm0.4$ & $ -4.49$ \\
$\log\varepsilon_\text{S}$  &              & $-6.3\pm0.5$ & $ -4.88$ \\
$\log\varepsilon_\text{Fe}$ & $-3.9\pm0.2$ & $-3.7\pm0.2$ & $ -4.50$ \\
$\log\varepsilon_\text{Ni}$ & $-5.0\pm0.1$ & $-3.7\pm0.3$ & $ -5.78$ \\
$v_\text{rad}$ [\kms] & $107.9\pm2.9$\tablefootmark{$\ast$} & $64.5\pm3.0$\tablefootmark{$\ast$} \\
$v_\text{rot}\,\sin i$ [\kms]   & $40\pm5$ & $25\pm5$\\
$d$ [pc]                & $508\pm17$    & $307\pm9$\\
$\dot M$ [\msr]         & $2.1\times10^{-9}$ & $<10^{-12}$\\
$v_\infty$ [\kms]       & 1570\\
$\dot M^\odot$ [\msr]         & $2.7\times10^{-9}$ & $3.9\times10^{-11}$\\
$v_\infty^\odot$ [\kms]       & 1550 & 1800\\
\hline
\end{tabular}\\
\tablefoot{Solar abundances were taken from \citet{asp09}. Blank items denote
values that were not determined.\\
\tablefoottext{$\ast$}{The radial velocity was derived from the UVES
spectrum.}}
\end{table}

The binary \hvezda\ (CD-44 2920, \radec{6}{48}{04.70}{-44}{18}{58.43}) consists
of a hot subdwarf and a compact companion \citep{thack}. The nature of the compact
object is unclear (see Sect.~\ref{evolim}). The orbital solution points to its
relatively high mass $M>1.2\,M_\odot$ \citep{mesci}. \hvezda\ is one of the few
hot subdwarfs that has been detected in the X-ray range \citep[see][for a review
on X-ray emission from hot subdwarfs]{sdxpreh}. Most of the  X-ray emission seen
from this binary is emitted by the compact companion of \hvezda, which must be
either a neutron star or a white dwarf, as evidenced by the  presence of a
significant periodicity at 13.2\,s in the X-ray emission
\citep{mesci,urychrot,popovchlazbt}. However, a significant X-ray flux with a
thermal spectrum and X-ray luminosity $3\times10^{30}\,\ergs$ is also clearly
detected when the compact companion is eclipsed by HD 49798, and can be
associated to the X-ray emission in the wind of the hot subdwarf itself
\citep{dvoj18}.

\subsection{Determination of stellar parameters}

\begin{table}[t]
\caption{Wavelengths of the strongest lines (in \AA) used for abundance
determination in \hvezda.}
\label{hd49798el}
\begin{tabular}{ll}
\hline
\ion{C}{iii}  & 4068, 4069, 4070\\
\ion{N}{iii, iv}  & 3999, 4004, 4058, 4097, 4103\, 4379, 4511, 4515, \\
              & 4518, 4524\\
\ion{O}{iii}  & 3757, 3760\,\\
\ion{Mg}{ii}  & 4481\\
\ion{Al}{iii} & 1855, 1863\\
\ion{Si}{iv}  & 3762, 3773, 4089, 4212, 4654\\
\hline
\end{tabular}
\end{table}

\begin{figure*}
\includegraphics[width=\textwidth]{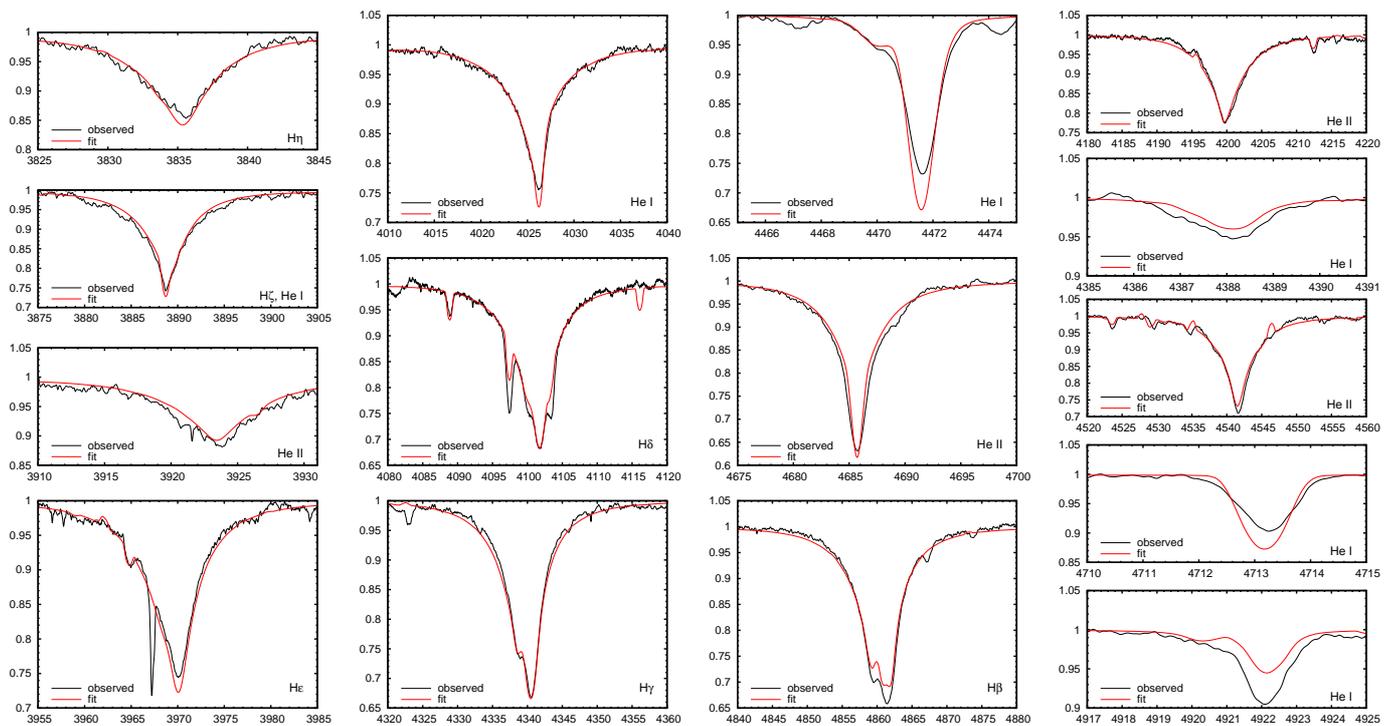}
\caption{Comparison of the best-fit synthetic spectra (red line) and UVES
spectra (black line) of \hvezda\ in the visual region. Here we plot the
normalised spectrum as a function of wavelength in \AA.}
\label{hd49798uves}
\end{figure*}

\begin{figure*}
\includegraphics[width=0.5\textwidth]{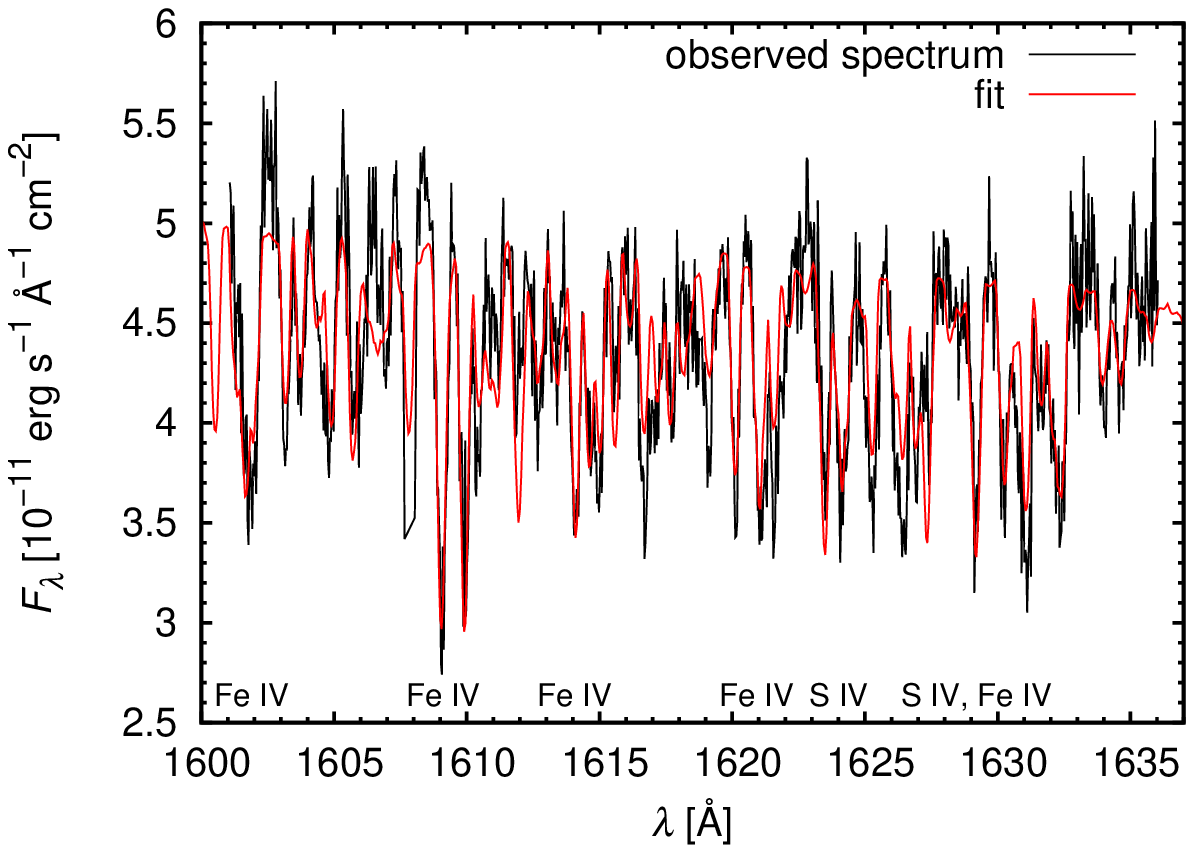}
\includegraphics[width=0.5\textwidth]{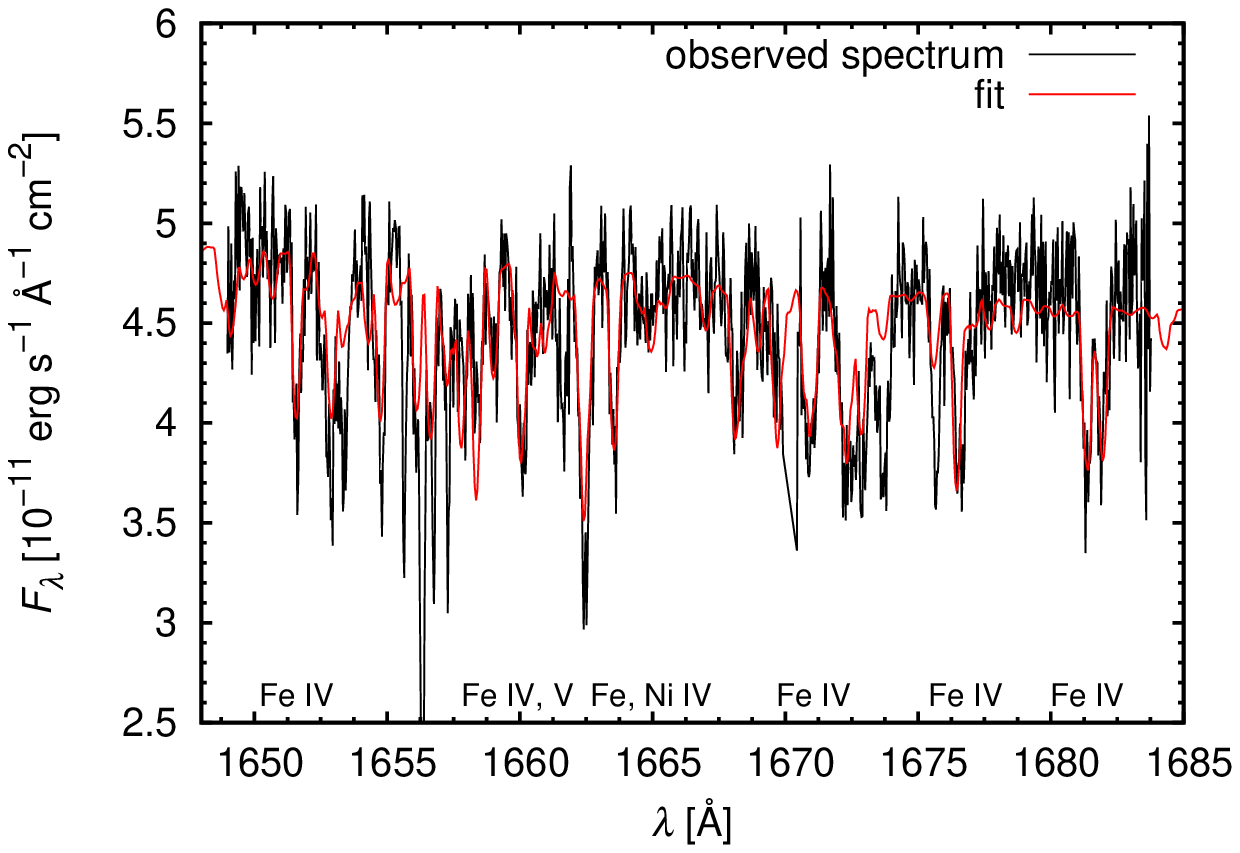}
\caption{Comparison of the best-fit synthetic spectra (red line) and HST/GHRS
spectra (black line) of \hvezda\ in the UV region.}
\label{hd49798hst}
\end{figure*}

We used UVES and HST/GHRS spectra for the determination of the stellar
parameters of \hvezda. All stellar parameters (see Table~\ref{hvezpar})
were derived from UVES spectra except for abundances of Al, Fe, and Ni, for
which we used HST/GHRS spectra. The model atmosphere grid for the stellar
parameter determination, $\Teff\in[44,\,46,\,48]\,\text{kK}$, $\log
(g/1\,\text{cm}\,\text{s}^{-2})\in[4.15,\,4.40,\,4.65,\,4.90]$, and
$\varepsilon_\text{He}\in[0.4,\,0.6,\,0.8]$, comprises the parameters derived by
\citet{dvoj16} $\Teff=47\,500\pm2000\,\text{K}$, $\logg=4.25\pm0.2$, and
$\varepsilon_\text{He}=0.5\pm0.1$. After the determination of the helium
abundance and the values of \Teff\  and $\log g$ from this grid, we determined
the abundances of other elements using a model with fixed $\varepsilon_\text{He}$,
\Teff,  and $\log g$. We iterated the process until convergence.

We adopted a slightly lower value of projected rotational velocity than
\citet{dvoj16} $v\sin i=40\,\kms$ that provides a better fit of metallic lines.
Figures~\ref{hd49798uves} and \ref{hd49798hst} compare the observed UVES and
HST/GHRS spectra with the best-fit synthetic spectra. The strongest lines used
for the abundance determination are summarised in Table~\ref{hd49798el}. We have
not listed the UV lines of Fe and Ni, which are too numerous. The derived
atmosphere parameters given in Table~\ref{hvezpar} agree with the results of
\citet{dvoj16} within errors except the derived surface gravity, which is
slightly higher. The nitrogen abundance is rather uncertain, because different
nitrogen lines give very different abundances. The derived chemical composition
deviates significantly from the solar composition (see Table~\ref{hvezpar}).
While carbon and oxygen are depleted, the abundances of nitrogen, iron, and
nickel are a factor of a few higher than the solar value \citep{asp09}.

The distance of \hvezda\ was derived using GAIA DR2 data \citep{gaia1,gaia2}.
With $V=8.287\pm0.003\,$mag \citep{lau} this gives the absolute magnitude
$M_V=-0.24\pm0.07\,$mag and with the bolometric correction $\text{BC} = 27.58 -
6.80 \log T_\text{eff}$ \citep{okali} the estimated luminosity is
$L=4400\pm400\,L_\odot$ \citep[Eq.~(5)]{okali}. From this latter value, we derive  a stellar radius of
$R=1.05\,R_\odot\pm0.06\,R_\odot$. With spectroscopic surface gravity, this
gives the mass $M=1.46\pm0.32\,M_\odot$, which nicely agrees with the mass of
$1.50\pm 0.05\,M_\odot$ derived from the orbital solution \citep{mesci}. With
$i=85^\circ$ and the radius, and $v_\text{rot}\sin i$ from Table~\ref{hvezpar},
we derive a  rotational period of $1.3\pm0.2\,$d. This is very close to the orbital
period 1.55\,d pointing to synchronised rotation.

\begin{figure}[t]
\centering
\resizebox{\hsize}{!}{\includegraphics{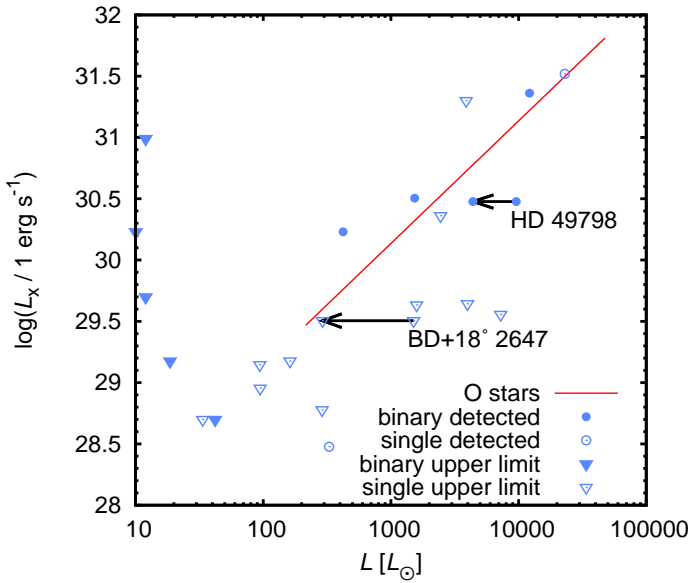}}

\caption{Relation between observed X-ray luminosity and the stellar
luminosity for subdwarfs (adopted from \citealt{snehurka}). Blue symbols refer
to individual subdwarfs: circles denote X-ray-detected subdwarfs, while
triangles denote available upper X-ray detection limits, filled symbols denote
subdwarfs in binaries, and empty symbols correspond to single objects.
Overplotted is the extrapolation of the observed mean relation for O stars
\citep[solid red line]{naze}. The shift of the stellar parameters with respect
to previous determinations is denoted using the black arrow.}

\label{lxlbolhdbd}
\end{figure}

The derived radius is lower than that estimated by \citet{dvoj16} due to lower
adopted distance and higher derived gravity. As a result, predicted luminosity
is also lower, which moves the position of the star closer to the canonical
relationship between the stellar luminosity and X-ray luminosity (see
Fig.~\ref{lxlbolhdbd}, which is plotted assuming intrinsic wind X-ray luminosity
of \hvezda\ $3\times10^{30}\,\ergs$, \citealt{dvoj18}). Consequently, in this
case the offset of the position of the star from the mean relationship was due
to imprecise stellar parameters.

\subsection{Wind model}
\label{hdvit}

\begin{figure}[t]
\centering
\resizebox{\hsize}{!}{\includegraphics{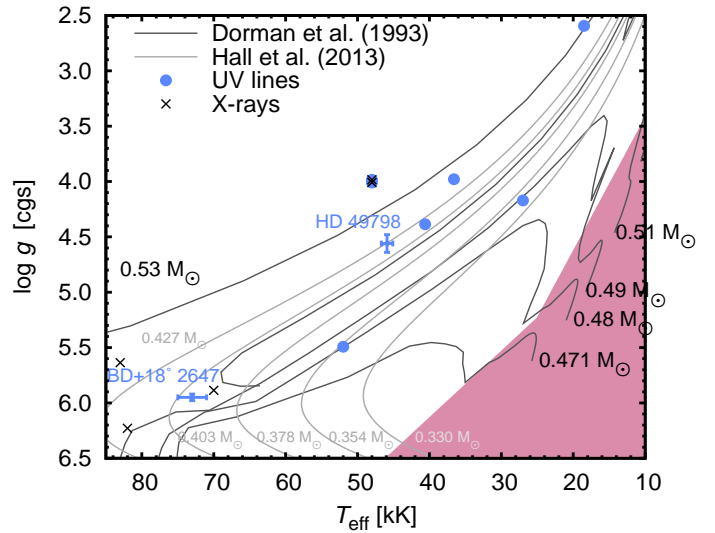}}
\caption{Position of studied stars in the \diag\ diagram. Red region denotes the
parameter area with no predicted wind \citep{snehurka}. Overplotted are the
evolutionary tracks of \citet{durman}, post-RGB tracks of \citet{hall8000} and
the positions of subdwarfs with known mass-loss rates derived from observed
UV wind-line profiles (blue circles) and subdwarfs with X-ray emission (black
crosses). Adapted from \citet{snehurka}.}
\label{sitpotoxhrvithdbd}
\end{figure}

We used our wind code to predict the structure of the stellar wind of \hvezda.
The subdwarf lies well above the wind boundary in the \diag\ diagram (see
Fig.~\ref{sitpotoxhrvithdbd}), and consequently the predicted mass-loss rate is
relatively large, $2.1\times10^{-9}\,\msr$ (see Table~\ref{hvezpar}). Although
iron and nickel are significantly overabundant with respect to the ratio of
abundances of these elements relative to hydrogen, the predicted wind mass-loss
rate is slightly lower than the value derived assuming solar chemical
composition, $2.7\times10^{-9}\,\msr$. This is because (for solar chemical
composition) the contribution of iron and nickel to the radiative force is
surpassed by that of oxygen, which is depleted on the surface of \hvezda\ by
more than one order of magnitude with respect to the solar value. Moreover, the
total mass fraction of heavier elements is close to the solar value. Similar
results were derived for the influence of CNO cycle abundances on mass-loss rate
for O stars \citep{dusik}.

The derived mass-loss rate is in close agreement with the value $2.8\times10^{-9}\,\msr$
derived from the fitting formula for subdwarfs \citep{snehurka} assuming
solar chemical composition. The mass-loss rates derived using the formula for ten
times higher and ten times lower abundances of heavier elements are
$8.8\times10^{-9}\,\msr$ and $8.7\times10^{-10}\,\msr$, respectively. The moderate
difference between these values and the predicted mass-loss rate
further demonstrates that the mass-loss rate is not particularly sensitive to
abundances for \hvezda\ parameters.

The X-ray luminosity of \hvezda\ outside eclipses stems from the release of the
gravitational potential energy during wind accretion on the compact companion.
Within the classical Bondi-Hoyle-Lyttleton theory \citep{holy,boho} the
accretion luminosity is
\begin{equation}
\label{lxlxrov}
\lx=\eta\frac{G^3\x M^3}{\x R D^2 v^4}\dot M,
\end{equation}
where $\x M$ and $\x R$ are the mass and radius of the compact companion, $D$ is
binary separation, $v$ is relative velocity, and $\eta$ is efficiency. If the
wind velocity is much larger than the orbital velocity and is not affected by X-ray
ionisation \citep{irchuch,sandvelax}, the wind terminal velocity can be
inserted instead of the relative velocity $v=v_\infty$. Using wind parameters
derived here (Table~\ref{hvezpar}) in conjunction with $\x M=1.28\,M_\odot$ and
$D=7.85\,R_\odot$ \citep{mesci}, with $\eta=1$ we derive 
$\lx=1.2\times10^{31}\,\text{erg}\,\text{s}^{-1}$ for typical parameters of
a white dwarf and $\lx=4\times10^{33}\,\text{erg}\,\text{s}^{-1}$ for a neutron
star.  The observed X-ray luminosity, with the updated distance provided by
GAIA, is $\sim10^{32}\,\ergs$ \citep{urychrot}. Consequently, to explain
this luminosity in the case of a white dwarf, a decrease of the wind velocity
due to X-ray ionisation is required. Instead, in the case of a neutron star, the
predicted luminosity is higher than the observed one, requiring a low
efficiency.

\subsection{The X-ray eclipse light curve}

\begin{figure}[t]
\centering
\resizebox{\hsize}{!}{\includegraphics{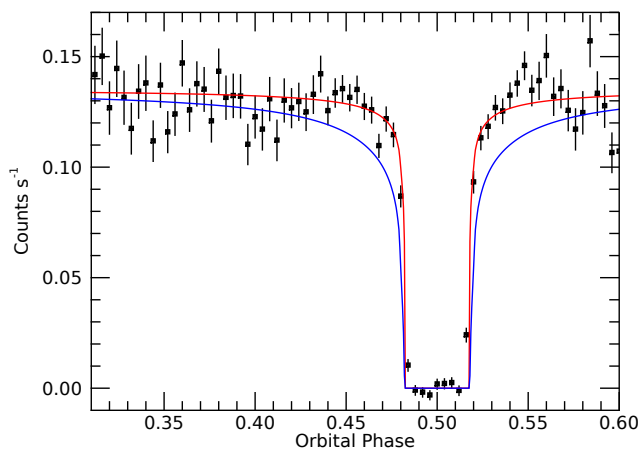}}

\caption{X-ray light curve of \hvezda\ plotted during the eclipse. The red line
shows the simulated eclipse light curve predicted assuming derived wind and
stellar parameters. For a comparison, we also plot the light curve calculated
assuming solar chemical composition (blue line).}

\label{wind_eclipse}
\end{figure}

X-ray observations with the XMM-Newton satellite showed that the eclipse ingress
and egress are not sharp, suggesting that the X-rays emitted by the compact
object are gradually absorbed in the wind of the sdO star. Using the wind model
and surface composition derived in this work, we computed the absorption
coefficient at different X-ray energies $\chi_\nu(r)$ and derived the expected
profile for the resulting light curve around the time of the eclipse
as an integral
\begin{equation}
F(\varphi)=F_0\,e^{-\tau(\varphi)},\qquad \tau(\varphi)=
\int_{z_\text{c}(\varphi)}^\infty\chi_\nu(r(z))\,\de z.
\end{equation}
Here, $F_0$ is the X-ray flux from the companion and the integral is calculated
numerically along the phase $\varphi$-dependent ray between the companion
(located at $z=z_\text{c}$) and observer. We assumed a circular orbit with an
inclination of $85^\circ$ that fits the length of the eclipse.

The calculation is compared to the observations in Fig.~\ref{wind_eclipse}. The
data were obtained from nine XMM-Newton observations covering the orbital
phase of the eclipse (2008 May 10; 2011 May 02; August 18, 20, 25; September 3,
8; 2013 November 9; 2018 November 8). The figure shows the net light curve in
the 0.15-0.5\,keV energy range as measured with the EPIC pn instrument (the
emission from \hvezda, as derived in the 4300\,s of complete eclipse, has been
subtracted).  The red line shows the expected profile of the eclipse as computed
with the wind abundances of HD 49798 derived in Sect.~\ref{hdvit}.  For
comparison, the blue line shows the  profile that would be caused by a stellar
wind with solar abundances. Although the statistical quality of the current
X-ray data does not allow us to directly estimate the wind parameters, it is clear
that the model derived with the proper abundances provides a good description of
the X-ray data.

\begin{table}[t]
\caption{Surface abundances of \hvezda\ expressed relative to the total number
density of baryons.}
\label{hdbar}
\centering
\begin{tabular}{ccc}
\hline
&\hvezda & Sun\\
\hline
$\log\tilde\varepsilon_\text{CNO}$ & $-3.7\pm0.2$ & $-3.21$\\
$\log\tilde\varepsilon_\text{Mg}$  & $-4.8\pm0.5$ & $-4.53$\\
$\log\tilde\varepsilon_\text{Al}$  & $-5.5\pm0.5$ & $-5.68$\\
$\log\tilde\varepsilon_\text{Si}$  & $-4.7\pm0.2$ & $-4.62$\\
$\log\tilde\varepsilon_\text{Fe}$  & $-4.5\pm0.2$ & $-4.63$\\
$\log\tilde\varepsilon_\text{Ni}$  & $-5.6\pm0.2$ & $-5.91$\\
\hline
\end{tabular}
\end{table}

\subsection{Evolutionary implications}
\label{evolim}

\citet{biss} propose that the progenitor of \hvezda\ was a star with an initial mass of
$4-6\,M_\odot$ which during the asymptotic giant branch phase lost its envelope
due to the common envelope event. This would imply that the star is in the phase
of shell-helium burning with a degenerate C-O core. However, stars with C-O
core mass implied by this scenario would instead ignite carbon
\citep{doktormasii,doktormasiii}. Moreover, massive post-AGB objects evolve on an
extremely fast evolutionary timescale of the order of years
\citep{vasiwo,milbe} and are therefore unlikely to be spotted in this stage.
Alternatively, massive subdwarfs could be formed by the merger of white dwarfs,
but this would likely lead to hydrogen deficient objects \citep{sajesplyn}.

\citet{popovchlazbt} suggested that the subdwarf originates from an  object with a helium-burning 
core. The companion could be either a white dwarf or a neutron star,
but the white dwarf is preferred on the basis of known properties of X-rays
and on evolutionary grounds \citep{biss}. Moreover, the detected spin-up
of the compact companion \citep{urychrot} could be naturally explained as the
consequence of cooling and contraction of a  white dwarf \citep{popovchlazbt}. 

\hvezda\ shows very unusual chemical composition (see Table~\ref{hvezpar}).
Chemical peculiarities in blue horizontal branch stars are typically attributed
to radiative diffusion \citep{huihb,miririhb,leblahb}. A mass-loss rate of
the order of $10^{-13}\,\msr$ is required to explain the observed abundance
anomalies in sdB stars \citep{unbuhb}. Higher mass-loss rates would not allow
for the abundance separation in the atmosphere, while weaker wind would lead to
a complete absence of helium \citep{unbuhb,ads}. Given the relatively high
mass-loss rate found in \hvezda, the detected abundance anomalies are
likely of evolutionary origin.

The high abundance of helium likely results from the stripping of the stellar
envelope which led to the exposition of the stellar core, whose chemical
composition was affected by hydrogen burning. A typical abundance ratio
resulting from hydrogen burning by CNO cycles is
$\log(\varepsilon_\text{O}/\varepsilon_\text{N})\approx-1$ \citep{biblerot},
while our analysis gives
$\log(\varepsilon_\text{O}/\varepsilon_\text{N})=-1.5\pm0.3$. The CNO
equilibrium carbon-to-nitrogen ratio is
$\log(\varepsilon_\text{C}/\varepsilon_\text{N})\approx-1.6$, which is
consistent with our result
$\log(\varepsilon_\text{C}/\varepsilon_\text{N})<-1.1$. To account for the enhanced
abundance of helium, we scaled the derived elemental abundances relative to
the baryonic number density $\tilde\varepsilon_\text{el}=
\varepsilon_\text{el}/(\varepsilon_\text{H}+4\varepsilon_\text{He})$. From
Table~\ref{hdbar} it follows that the scaled abundances of most elements are
consistent with solar chemical composition.

According to the evolutionary models, the helium-shell-burning subdwarf will
fill its Roche lobe  in approximately 40\,000 -- 65\,000\,yr
\citep{jared,wangnebang}. On this evolutionary time-scale, the current
mass loss as predicted here will not significantly contribute to mass transfer.
The subdwarf lifetime is of the order of 1 Myr if the subdwarf is indeed 
burning its core helium \citep{pacyhe}, implying a larger contribution of the wind to the mass
transfer. \citet{mesci} showed that the further evolution of the system could
lead to a second phase of mass transfer, which could ignite a SN~Ia explosion.
An alternative outcome of the evolution could be a double degenerate binary
consisting of a neutron star (originating from the collapse of the compact
companion) and a white dwarf \citep{jared,wangnebang}.
 
\section{BD+18$^\circ$\,2647}
\renewcommand\hvezda{\object{BD+18$^\circ$\,2647}}

Subdwarf \hvezda\ (Feige 67, PG 1239+178, \radec{12}{41}{51.79}{+17}{31}{19.75})
is a helium-poor star, which does not show any signs of a secondary companion
\citep{latour4}. The subdwarf does not show any detectable X-ray emission with
the upper limit of X-ray luminosity $3.2\times10^{29}\,\ergs$ \citep{bufacek}.

\subsection{Determination of stellar parameters}

\begin{table}[t]
\caption{Wavelengths of the strongest lines (in \AA) used for abundance
determination in \hvezda.}
\label{bd182647el}
\begin{tabular}{ll}
\hline
\ion{C}{iii}  & 977\\
\ion{C}{iv}   & 1169, 1548, 1551\\
\ion{N}{iv}   & (923), (924), (955), 1719\\
\ion{N}{v}    & 1239, 1243 \\
\ion{O}{iv}   & 923\\
\ion{F}{iv}   & 1060\\
\ion{Ne}{v}   & 1721\\
\ion{Si}{iv}  & 1067, 1122, 1128, 1394, 1403 \\
\ion{S}{v}    & 1040, 1502, 1572\\
\ion{S}{vi}   & 933, 944, 1001, 1118\\
\ion{Fe}{v}   & 1068, 1200--1208, 1234--1799\\
\ion{Fe}{vi}  & 986, 1009, 1042, 1200--1210, 1228--1766\\
\ion{Ni}{v}   & 1001, 1200--1211, 1225--1620\\
\ion{Ni}{vi}  & 954, 971, 980, 982, 1000, 1003, 1009, 1042, 1061, \\
              & 1065, 1068, 1070, 1071, 1075, 1076, 1200--1211,\\
              & 1227--1620\\
\hline
\end{tabular}
\tablefoot{Lines of \ion{N}{iv} given in parentheses, which point to a higher
nitrogen abundance than given in Table~\ref{hvezpar}, were heavily contaminated
by interstellar lines and were not used to determine abundances.}
\end{table}

\begin{figure*}
\includegraphics[width=\textwidth]{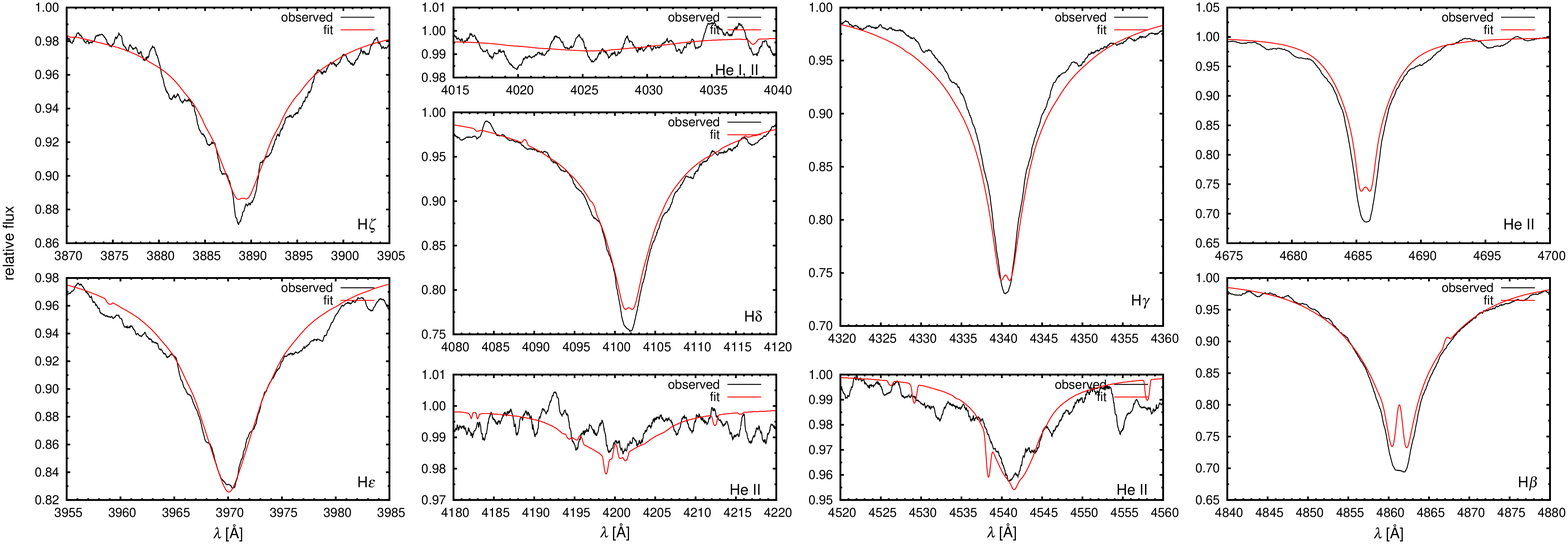}
\caption{Comparison of synthetic spectra calculated for the derived parameters
and UVES spectra of \hvezda\ in the visual region.}
\label{BD182647uves}
\end{figure*}

\begin{figure*}
\includegraphics[width=\textwidth]{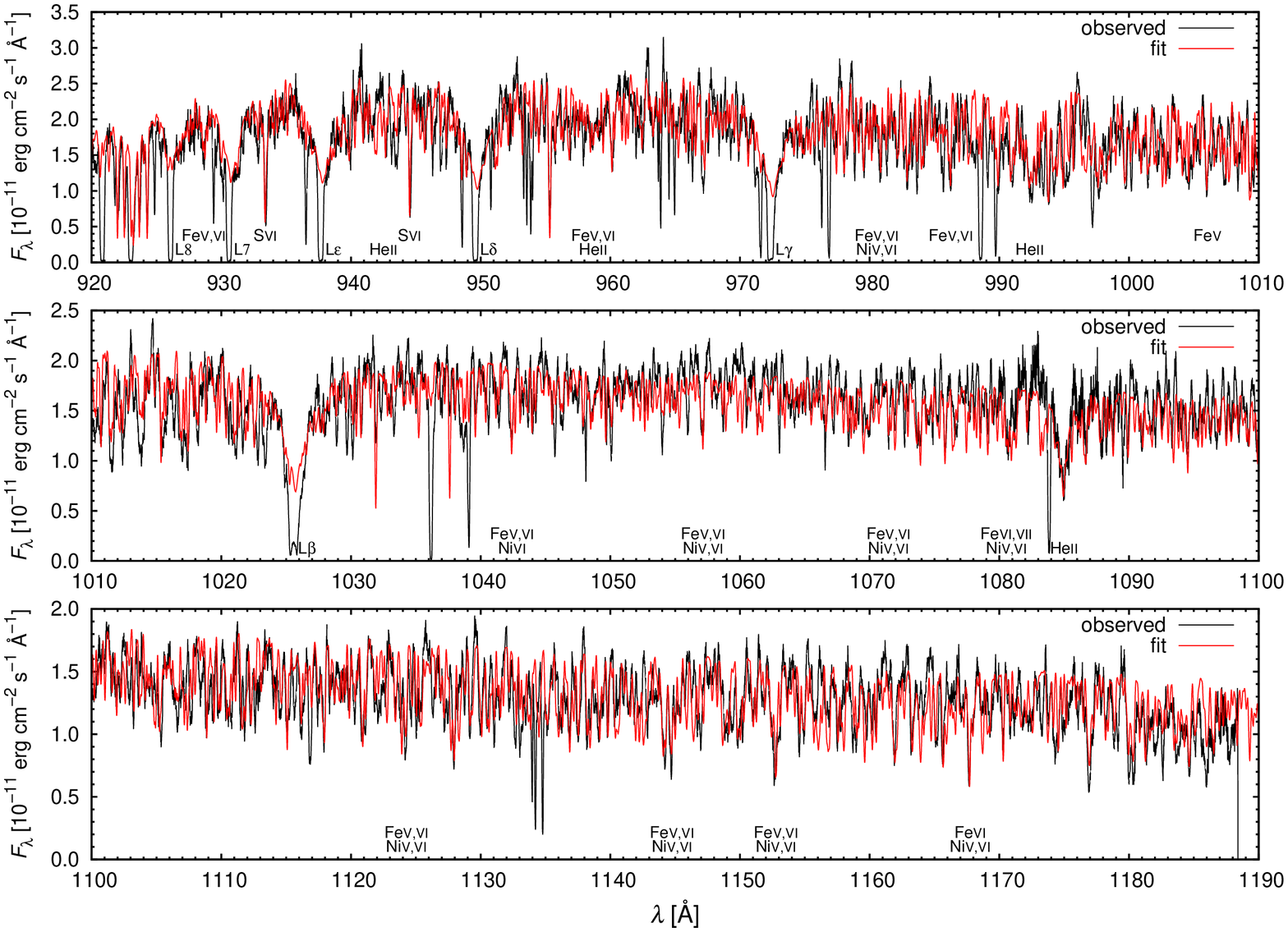}
\caption{Comparison of synthetic spectra calculated for the derived
parameters and FUSE spectra of \hvezda\ in the UV region.}
\label{hd49798fuse2}
\end{figure*}

We used UVES, IUE, and FUSE  spectra for the determination of the stellar
parameters of \hvezda.
The $T_\text{eff}$, $\log g$, and helium abundance were determined from the fit of
optical and Lyman lines. The cores of predicted optical lines show emission and
the cores of Lyman lines are affected by interstellar absorption. Therefore we
fitted only the wings in the case of hydrogen Lyman lines and Balmer lines with central 
emission. The wings originate in the lower parts of the
photosphere and are not severely affected by NLTE effects.
For the determination of \hvezda\ parameters, we first
used a grid of NLTE models with a combination of $\Teff\in[
55,\,60,\,65,\,70,\,75,\,80,\,85,\,90]\,$kK, $\log
(g/1\,\text{cm}\,\text{s}^{-2})\in[4.5,\,5.0,\,5.5,\,6.0]$, and
$\varepsilon_\text{He}\in[0.01,\,0.032,\,0.1]$. During the final steps of
parameter determination, we restricted the grid to
$\Teff\in[65,\,70,\,75]\,\text{kK}$, $\log
(g/1\,\text{cm}\,\text{s}^{-2})\in[5.5,\,6.0,\,6.5]$, and
$\varepsilon_\text{He}\in[0.01,\,0.032,\,0.1]$. Using these grids we obtained
the values of \Teff, $\log g$, and $\varepsilon_\text{He}$ listed in
Table~\ref{hvezpar}. The abundances of heavy elements were determined from the
fit of UV spectra with a model calculated for derived values of stellar
parameters. These steps were repeated until convergence was achieved.

The strongest lines used for abundance determination are given in
Table~\ref{bd182647el}. The comparison of the observed UVES and FUSE spectra and
the best-fit synthetic spectra is given in Figs.~\ref{BD182647uves} and
\ref{hd49798fuse2}. The derived atmosphere parameters and abundances given in
Table~\ref{hvezpar} agree with the results of \citet{latour4} within errors,
except for the derived effective temperature which is higher by about 12~kK.
\citet{latour4} used optical spectra with lower resolution, consequently the
problems with line centre emission could be one of the reasons for the
difference in the derived temperatures. We were able to obtain a reasonable fit
of hydrogen lines for lower $T_\text{eff}$, but this led to overly strong
\ion{He}{i} lines. Our derived effective temperature is lower than
$T_\text{eff}=75\,000\,\text{K}$ derived by \citet{sam5}.

To test the origin of the difference between the effective temperature
derived by \citet{latour4} and our results, we smoothed the observed spectra by
a Gaussian filter with $1.3\,\AA$ width that should roughly correspond to the
spectra used in their analysis. We fitted full line profiles and the derived
effective temperature was lower by about $6\,$kK. This shows that the adopted
spectral resolution in combination with a different way of fitting could be one
of the causes of the difference between the results. A higher abundance of iron, as
derived by \citet{latour4}, could be another reason for the difference. For some
stars, the use of UV and optical observations may lead to discrepant results
\citep{dichala}, however in our case the UV spectra helped us mainly to
constrain the grid of possible stellar parameters.

\begin{figure}
\includegraphics[width=0.5\textwidth]{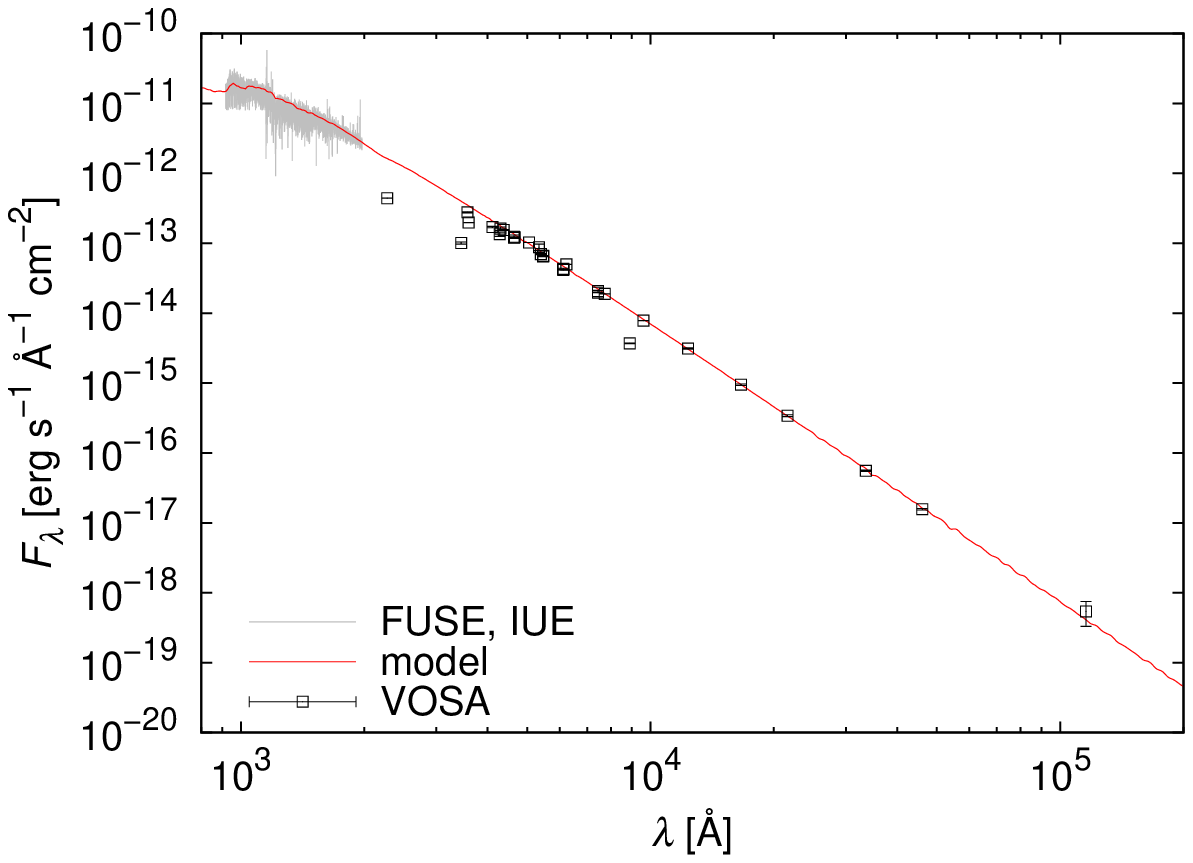}
\caption{Comparison of predicted (red line) and observed flux distribution. The
predicted flux was taken from the best-fit model and smoothed using a Gaussian
filter. The observed flux distribution was taken from FUSE and IUE observations
(grey lines) and from the optical and inferred photometry available in the VOSA
database (squares with error bars).}
\label{BD182647sed}
\end{figure}

Because the derived effective temperature is higher than that derived by
\citet{latour4}, we further compared the predicted flux distribution with that
derived using the VOSA tool \citep{vosa}. The observed flux distribution in
Fig.~\ref{BD182647sed}  is in close agreement with predictions with $E(B-V)=0.04\pm0.02$
and assuming the \citet{fima} extinction law. The derived reddening agrees with
$E(B-V)=0.02\pm0.02$ \citep{prasan}, which corresponds to \hvezda\ coordinates.

The distance was derived using GAIA DR2 data \citep{gaia1,gaia2}. With
$V=11.63\pm0.12\,$mag \citep{hogefabe} this gives the absolute magnitude
$M_V=4.07\pm0.15\,$mag (with $E(B-V)=0.04\pm0.02$) and with the bolometric
correction $\text{BC} = 27.58 - 6.80 \log T_\text{eff}$ \citep{okali} the
estimated luminosity is $L=290\pm50\,L_\odot$ \citep[Eq.~(5)]{okali}. From this,
the stellar radius is $R=0.107\pm0.011\,R_\odot$. With spectroscopic surface
gravity this gives a mass of $M=0.38\pm0.08\,M_\odot$.

The adopted bolometric correction fit \citep{okali} was derived for O stars
with significantly lower effective temperature ($T_\text{eff}<45\,$kK) than
derived here. To estimate the error connected with extrapolation of the
bolometric correction, we calculated bolometric correction from our models using
Eq.~(1) of \citet{ostar2003} and the$V$ filter response curve from Mikul\'a\v sek
(private communication). The derived $BC=-5.60\pm0.08\,$mag is only slightly
lower than estimated from the fit of \citet{okali}, and therefore the resulting
stellar parameters ($L=320\pm50\,L_\odot$, $R=0.113\pm0.011\,R_\odot$, and
$M=0.41\pm0.09\,M_\odot$) only differ within the derived uncertainties. We also
note that the application of the frequently used bolometric corrections of
\citet[see also \citealt{spravnybyk},]{bckytka} to \hvezda\ also leads to
extrapolation, because these corrections were derived for stars with
$T_\text{eff}<53\,$kK. Furthermore, \citet{bckytka} used high-order
polynomial approximation, which may give erroneous results during extrapolation.
In our case, the formula predicts bolometric correction which is 1\,mag higher
than adopted here leading to significantly lower mass.

The central parts of predicted hydrogen line profiles show emission that is
missing in observed spectra (see Fig.~\ref{BD182647uves}). The missing emission
in the observed spectra was already recognised by \citet{latour4}, but the
absence of emission is more striking in our analysis because we use spectra with
higher resolution. To understand the origin of the emission in theoretical
spectra, we calculated an additional NLTE model with fixed temperature derived
assuming LTE. Even this model predicts emission in hydrogen line profiles.
Therefore, the emission in predicted spectra is not caused by NLTE temperature
inversion.

On the other hand, the structure of model atmosphere could be influenced by
elements not included in our analysis. This could be connected with the possible
appearance of weak lines in the optical spectrum which is not explained by the
predicted spectrum. To test this, we calculated additional model
atmospheres with ten times higher abundance of selected elements whose
abundances were not determined from spectroscopy (Mg, Al, P, and Cr). However,
we did not find any significant influence of these elements on hydrogen line
profiles. Moreover, classical chemically peculiar stars show vertical abundance
stratification \citep{tuhtalestra,nesvacil}, which may be present also in
subdwarf atmospheres and affect the observed spectra (as suggested, e.g. by
\citealt{geimet}).

\subsection{Wind model}

Also, \hvezda\ is located outside the wind limit in \diag\ diagram
(Fig.~\ref{sitpotoxhrvithdbd}, see also \citealt{snehurka}), however the
mass-loss rate $7.2\times10^{-11}\,\msr$  predicted using Eq.~(1) of
\citet{snehurka} for \hvezda\ parameters (Table~\ref{hvezpar}) and solar
chemical composition is two orders of magnitude lower than that of HD\,49798.
This value agrees reasonably well with the mass-loss rate of $3.9\times10^{-11}\,\msr$
derived from global models and solar chemical composition.

Due to its low mass-loss rate, the solar abundance wind of \hvezda\ is mostly
driven by lighter elements O, Ne, and Mg, while the contribution of iron is
relatively small because numerous iron lines remain optically thin at low
density \citep[e.g.][]{pusle,vinca,snehurka}. Because these lighter
elements are depleted on the surface of \hvezda\  (see Table~\ref{hvezpar}), it can be
expected that the mass-loss rate calculated for realistic \hvezda\ abundances is
significantly lower than the mass-loss rate derived using solar abundances.

To test the presence of the wind for \hvezda\ abundances we compared the
magnitude of the radiative acceleration due to lines and scattering on free
electrons $g^\text{rad}$ with the magnitude of the gravitational acceleration
$g$. The winds are only possible if 
\begin{equation}
\label{trebova}
g^\text{rad}>g.
\end{equation}
The wind condition of Eq.~\eqref{trebova} was tested using artificial wind
models with fixed density and velocity structure and with mass-loss rate as a
parameter. We assumed a linear velocity profile and the density profile was
derived from the assumed mass-loss rate using the continuity equation. The
application of Eq.~\eqref{trebova} to these artificial models led to
an upper limit of the mass-loss rate of  \hvezda\  of $10^{-12}\,\msr$, while the wind
condition is fulfilled for a mass-loss rate of $10^{-13}\,\msr$.

Nevertheless, Eq.~\eqref{trebova} gives a simply necessary but not sufficient condition
for the presence of wind, because it does not compare the radiative and
gravitational acceleration for a consistent wind solution. This likely explains
why our global models failed to provide consistent wind models, although
Eq.~\eqref{trebova} would allow for a wind, albeit with a low mass-loss rate. To
better understand this issue, we calculated additional simplified (but consistent)
wind models that use the model atmosphere flux as an input for the calculation
of the radiative force. Our tests with these simplified models showed that they
are unable to pass through the sonic point with consistently calculated
radiative force. Close inspection of the results revealed that the models are
able to provide a converged solution with a first estimate of the radiative force,
but further iterations failed. This corresponds to the fact that the models
fulfilled the necessary condition for the presence of a wind Eq.~\eqref{trebova}
(which uses just one calculation of the radiative force), but failed to provide
a consistent model.

Therefore, it is likely that \hvezda\ is located close to the wind limit for its
abundances. The proximity of the wind limit implies that  the
wind mass-loss rate is sensitive to metallicity. This can be further demonstrated using
the mass-loss rate formula of \citet{snehurka}, which for the increase of
abundance by a factor of ten from the solar values predicts an increase of the
mass-loss rate also by a factor of ten. This is a stronger metallicity
dependence than in HD~49798. We cannot completely rule out the presence of
wind in \hvezda\ because the abundances of some elements that can drive
wind were not determined from the spectra and could be higher than the solar values
assumed here. Moreover, stars close to the wind limit may still have very weak
purely metallic wind \citep{babela}. The missing (or very weak) wind explains
the absence of X-rays in \hvezda\
\citep[$\lx<3.2\times10^{29}\,\ergs$,][]{bufacek}.

\subsection{Evolutionary implications}

Hot subdwarfs are typically expected to be the products of binary evolution
\citep{han}. However, understanding the evolutionary state of \hvezda\ is
complicated by the unknown nature of a potential secondary. There is no evidence of
a secondary from radial velocities, spectral energy distributions
\citep[see also Fig.~\ref{BD182647sed}]{wiltom,latour4}, or from optical
variability \citep{standardgaia2,standardgaia}. This could possibly mean that
either the secondary is on a wide orbit, or that it is missing. Another problem
is that the mass determined here is relatively small compared to typical
subdwarfs \citep{heberpreh}, although the derived uncertainties are large and in
principle even allow for more typical subdwarf masses.

From the missing signature of a companion it follows that \hvezda\ could be formed
by stable Roche-lobe overflow near the tip of the red giant branch, which may form
wide binaries with low-mass subdwarfs \citep{podhanlyn,vospermas}. Alternatively,  \hvezda\
could have been produced by the merger of either two helium white dwarfs \citep{han} or a
helium white dwarf and a main sequence star \citep{zhanghehp}. However, the
merger scenario likely implies a He-rich composition, which is not the case for
this star. \hvezda\ parameters also nicely correspond to stars in a post-red-giant-branch evolutionary phase \citep[Fig.~\ref{sitpotoxhrvithdbd},][see
\citealt{reindlobr}, Fig.~13]{hall8000}, which have a degenerate helium core
formed as a result of common envelope evolution. However, a companion on
a short-period orbit is expected in this case. On the other hand, the derived mass
is still within uncertainties close to the canonical subdwarf mass, which would
allow for evolutionary scenarios that are more typical for subdwarfs
\citep{heberpreh}.

The derived peculiar composition (Table~\ref{hvezpar}) likely originates from
radiative diffusion and gravitational settling in the atmosphere of the star
\citep{latour4}. The upper limit of the mass-loss rate for  \hvezda\  has just the right
value to moderate abundance stratification \citep{unbu98,unbuhb}, while stronger
wind would inhibit the abundance stratification \citep{unbuhb,vimiri}. Within
diffusion models, low abundance of helium is a result of gravitational settling.
\citet{chadif} provided equilibrium abundances of white dwarfs accounting for
radiatively supported diffusion and, in agreement with our results, found
enhanced surface abundance of iron, while carbon, nitrogen, neon, silicon, and
sulphur were depleted at $\Teff=70\,\text{kK}$.

It is not clear how is it possible that the star acquired abundance anomalies
that would not likely develop if the star appeared at its current position on the HR
diagram with solar abundances. Likely, the star initially had solar chemical
composition, was located below the wind limit at some previous stage of its
evolution, and developed abundance anomalies by  diffusion. The
abundance anomalies then persisted until the star attained the current
parameters, although the wind predicted assuming a solar composition would
have wiped out any peculiarities \citep{unbuhb}.  It is also not clear why
helium is still present in the atmosphere, because with absence of a wind, helium
should be missing \citep{unbuhb}. Perhaps, some other process (e.g.
turbulence, \citealt{mirivi}) is responsible for this.

Despite similar abundance anomalies, in contrast to classical chemically
peculiar stars, the subdwarfs and horizontal branch stars do not show
rotationally modulated light variability
\citep{standardgaia2,standardgaia,ernstjakomy}. Light variability in classical
chemically peculiar stars typically originates due to flux redistribution in
surface abundance spots \citep[e.g.][]{prvalis}, which are supposed to be
connected with strong surface magnetic field. Hot subdwarfs likely do not
possess any strong global magnetic fields \citep{podland}, which might explain
the lack of rotationally modulated light variability.

\section{Conclusions}

We studied the implications of realistic surface abundances for the hot subdwarf
wind mass-loss rates. We determined stellar parameters for two selected
hydrogen-rich, hot subdwarfs from our own optical spectroscopy and from UV
spectroscopy using TLUSTY NLTE atmosphere models. We predicted the wind
mass-loss rates with our global wind model and compared the derived results with
those predicted assuming solar chemical composition.

For HD\,49798, we find an effective temperature in agreement with previous
determinations and a mass that agrees with a binary solution. The chemical
composition, with enhanced abundances of helium and nitrogen, appears to be a
result of a previous process of hydrogen burning. The mass-loss rate predicted
using realistic surface abundances does not significantly differ from that
derived using solar abundances, because the mass fraction of heavier elements
roughly corresponds to the solar chemical composition. The X-ray eclipse light
curve can be nicely reproduced by absorption in the wind with the derived
mass-loss rate and abundances, but not by a wind with solar abundances.

In the case of BD+18$^\circ$\,2647 the derived stellar parameters
agree reasonably with previous determinations with the exception of the effective
temperature, which is about 12\,kK higher than the recent determination, but
is still in the range of values available in the literature. The discrepant
temperature is probably due to an overly weak dependence of line profiles on the stellar
parameters and the appearance of emission in the cores of predicted absorption
lines. The subsolar abundance of light elements and the overabundance of iron can be
interpreted as a result of radiative diffusion and gravitational settling. As a
result of this, the homogeneous wind is missing for determined abundances, while
the wind would exist at solar metallicity. On the other hand, a stronger wind
would likely have effaced  all peculiarities.

Although we used UV and optical spectra to determine the abundances, the
lines of some elements that are important for  driving wind are inaccessible in
the available spectral regions. From our models it follows that about one-third of the line
driving still comes from such elements. This contributes to the uncertainty related to
mass-loss rate predictions.

We conclude that while the precise abundances are not very important for the
strength of the wind in cases where the surface abundances are affected by
hydrogen burning, abundances have a significant effect when diffusion
processes come into play. Therefore, abundance variations and imprecise stellar
parameters may be one of the reasons for the large scatter of hot subdwarf X-ray
luminosity when plotted as a function of bolometric luminosity.

\begin{acknowledgements}
This research was supported by grant GA\,\v{C}R 18-05665S. SM acknowledges
financial contribution from the agreement ASI-INAF n.2017-14-H.0. This project
has received funding from the European Union’s Framework Programme for Research
and Innovation Horizon 2020 (2014-2020) under the Marie Skłodowska-Curie grant
Agreement No. 823734. Computational resources were provided by the CESNET
LM2015042 and the CERIT Scientific Cloud LM2015085, provided under the programme
"Projects of Large Research, Development, and Innovations Infrastructures". The
Astronomical Institute Ond\v{r}ejov is supported by the project RVO:67985815.
This publication makes use of VOSA, developed under the Spanish Virtual
Observatory project supported from the Spanish MINECO through grant
AyA2017-84089.
\end{acknowledgements}

\bibliographystyle{aa}
\bibliography{papers}

\end{document}